\documentclass[12pt,preprint]{aastex}
\shorttitle{Infrared Observation toward HB21}
\shortauthors{Shinn et al.}

\newcommand{\luerg}{erg s$^{-1}$ cm$^{-2}$ sr$^{-1}$}
\newcommand{\ncm}{cm$^{-3}$}
\newcommand{\ncmK}{cm$^{-3}$ K}
\newcommand{\Ncm}{cm$^{-2}$}
\newcommand{\kms}{km s$^{-1}$}
\newcommand{\um}{$\mu$m}

\newcommand{\Hi}{H {\small I}}

\newcommand{\hh}{H$_2$}
\newcommand{\HH}{H$_2$ $\upsilon=1\rightarrow0$ S(1)}

\newcommand{\Neii}{[Ne {\small II}]}
\newcommand{\Neiii}{[Ne {\small III}]}

\newcommand{\CO}{$^{12}$CO $J=2\rightarrow1$}

\newcommand{\ionNeii}{Ne$^{+}$}
\newcommand{\ionNeiii}{Ne$^{2+}$}

\newcommand{\Xco}{$X_{\textrm{\tiny{CO}}}$}

\newcommand{\nhh}{$n(\textrm{H}_2)$}
\newcommand{\nh}{$n(\textrm{H})$}
\newcommand{\nH}{$n_{\textrm{\tiny{H}}}$}
\newcommand{\nhe}{$n(\textrm{He})$}
\newcommand{\vs}{$\upsilon_s$}

\newcommand{\Nhh}{$N(\textrm{H}_2)$}
\newcommand{\NhhIRC}{$N$(H$_{2}\,; T>100\,\textrm{K})$}
\newcommand{\Nco}{$N(\textrm{CO})$}

\newcommand{\akari}{\textit{AKARI}}
\newcommand{\spitzer}{\textit{Spitzer}}

\begin{document}

\title{Infrared Studies of Molecular Shocks in the Supernova Remnant HB21:\\ I. Thermal Admixture of Shocked \hh{} Gas in the North}

%\author{Jong-Ho Shinn et al.}
\author{Jong-Ho Shinn\altaffilmark{1}, Bon-Chul Koo\altaffilmark{1}, Michael G. Burton\altaffilmark{2}, Ho-Gyu Lee\altaffilmark{1,3}, Dae-Sik Moon\altaffilmark{4}}

\email{jhshinn@snu.ac.kr}
\altaffiltext{1}{Dept. of Physics and Astronomy, FPRD, Seoul National University, 599 Gwanangno, Gwanak-gu, Seoul, 151-747, South Korea}
\altaffiltext{2}{School of Physics, University of New South Wales, Sydney, New South Wales 2052, Australia}
\altaffiltext{3}{Astrophysical Research Center for the Structure and Evolution of the Cosmos, Sejong University, Seoul 143-747, Korea}
\altaffiltext{4}{Dept. of Astronomy and Astrophysics, University of Toronto, Toronto, ON M5S 3H4, Canada}

\begin{abstract}
We present near- and mid-infrared observations on the shock-cloud interaction region in the northern part of the supernova remnant HB21, performed with the InfraRed Camera (IRC) aboard \akari{} satellite and the Wide InfraRed Camera (WIRC) at the Palomar 5 m telescope.
The IRC 7 \um{} (S7), 11 \um{} (S11), and 15 \um{} (L15) band images and the WIRC \HH{} 2.12 \um{} image show similar shock-cloud interaction features.
We chose three representative regions, and analyzed their IRC emissions through comparison with \hh{} line emissions of several shock models.
The IRC colors are well explained by the thermal admixture model of \hh{} gas---whose infinitesimal \hh{} column density has a power-law relation with the temperature $T$, d$N\sim T^{-b}dT$---with \nhh{} $\sim10^3$ \ncm, $b\sim3$, and \NhhIRC{} $\sim3\times10^{20}$ \Ncm.
The derived $b$ value may be understood by a bow shock picture, whose shape is cycloidal (cuspy) rather than paraboloidal.
However, this picture raises another issue that the bow shocks must reside within $\sim0.01$ pc size-scale, smaller than the theoretically expected.
Instead, we conjectured a shocked clumpy interstellar medium picture, which may avoid the size-scale issue while explaining the similar model parameters.
The observed \HH{} intensities are a factor of $\sim17-33$ greater than the prediction from the power-law admixture model.
This excess may be attributed to either an extra component of hot \hh{} gas or to the effects of collisions with hydrogen atoms, omitted in our power-law admixture model, both of which would increase the population in the $\upsilon=1$ level of \hh.
%This excess may be attributed to either an extra component of hot \hh{} gas or to the omission of collisions with hydrogen atoms in the power-law admixture model, both of which would increase the population in the $\upsilon=1$ level of \hh.
\end{abstract}

\keywords{ISM: individual (SNR G89.0+4.7) --- (ISM:) supernova remnants --- ISM: clouds --- infrared: ISM --- turbulence --- shock waves}

\section{Introduction} \label{intro}
Shock-cloud interactions in supernova remnants (SNRs) have been studied for their wide astrophysical importance, such as the followings.
Firstly, we can study the physics and chemistry of molecular shocks \citep{Draine(1983)ApJ_264_485,Hollenbach(1989)ApJ_342_306}, to be used as a basis for interpreting astrophysical phenomena.
Secondly, the evolution of shocked molecular clouds is closely related to the star formation occurring in them \citep{Elmegreen(1978)ApJ_220_1051,Vrba(1987)ApJ_317_207,Reynoso(2001)AJ_121_347}.
Lastly, the evolution of the SNRs interacting with molecular clouds is presumed to be significantly different from that of SNRs that do not interact with molecular clouds \citep[e.g.][]{Rho(1998)ApJ_503_L167,Shinn(2007)ApJ_670_1132}.
In addition, SNR shocks may have a simpler configuration in their interaction region than do shocks in outflows, since the latter can be generated by a diverse range of bullets, clumps, and/or winds \citep[cf.~Fig.~11 in][]{Hollenbach(1989)inproc}. 
Hence, SNRs provide geometrically simpler conditions to study the shock-cloud interaction.

The shock-cloud interaction manifests itself through radiative means \citep{Draine(1993)ARA&A_31_373,Wardle(2002)Science_296_2350}: high-velocity wings on low-$J$ CO rotational lines \cite[e.g.][]{Koo(1997)ApJ_485_263,Koo(2001)ApJ_552_175}; high-$J$ CO rotational lines and OH rotational lines in the far-infrared \cite[e.g.][]{Watson(1985)ApJ_298_316,Melnick(1987)ApJ_321_530}; radio maser lines of OH \cite[e.g.][]{Frail(1994)ApJ_424_L111}; and infrared emission lines of \hh{} \cite[e.g.][]{Neufeld(2007)ApJ_664_890}.
Among these, the infrared emission lines of \hh{} are useful in studying the excitation condition of shocks, since \hh{} is the most abundant molecule in interstellar clouds \citep{Snow(2006)ARA&A_44_367} and the lines, originating from quadrupole transition, are optically thin and closely spaced through $\sim1-30$ \um.
In addition, high resolution imaging is easier for \hh{} infrared emission lines, than for radio emission lines.

Infrared \hh{} emission lines have been frequently observed in SNRs interacting with nearby clouds.
One noticeable common result is that the \hh{} level-population diagram shows an \emph{ankle-like} curve (cf.~Fig.~\ref{fig-pop}) in the range of energy levels, $0-25,000$ K (\citealt{Oliva(1990)A&A_240_453,Richter(1995)ApJ_454_277,Richter(1995)ApJ_449_L83,Rho(2001)ApJ_547_885,Neufeld(2008)ApJ_678_974}; see Fig.~7a in \citealt{Rho(2001)ApJ_547_885} for the \emph{ankle-like} curve), which has been usually interpreted as a population of two-temperature ($\sim10^2$ and $\sim10^3$ K) \hh{} gas in Local Thermodynamic Equilibrium (LTE).
Such a population cannot be explained by a single planar shock model \citep[e.g.~Table~2 of][]{Wilgenbus(2000)A&A_356_1010}.
Hence, several mechanisms have been proposed in order to explain this level population distribution, such as a partially dissociative J-shock \citep{Brand(1988)ApJ_334_L103,Burton(1989)inproca,Moorhouse(1991)MNRAS_253_662}, a bow shock \citep{Smith(1991)MNRAS_248_451}, and a non-stationary shock \citep{Chieze(1998)MNRAS_295_672,Cesarsky(1999)A&A_348_945}.

However, the interpretation of the \hh{} population diagram is still under debate; for instance, even for the same target SNR, IC443, all the above three models were preferred by different authors \citep{Richter(1995)ApJ_454_277,Cesarsky(1999)A&A_348_945,Neufeld(2008)ApJ_678_974}.
One restriction that hinders a clear understanding of the \hh{} population diagrams, and subsequently of the shock-cloud interaction, is that the observations hitherto performed were not able to resolve the interaction features sufficiently to distinguish their structures, such as planar shocks, bow shocks, shocked clumps etc.
Our target SNR, G89.0+4.7 (conventionally known as HB21), provides an excellent example to study the shock-cloud interaction, avoiding this restriction, because of its large angular size and proximity.

HB21 is a large ($\sim120\arcmin\times90\arcmin$), middle-aged ($\sim5000-7000$ yr, \citealt{Lazendic(2006)ApJ_647_350,Byun(2006)ApJ_637_283}) SNR, located at the distance of $\sim0.8-1.7$ kpc \citep{Leahy(1987)MNRAS_228_907,Tatematsu(1990)A&A_237_189,Byun(2006)ApJ_637_283}.
It has been suspected of interacting with a molecular cloud, because of its deformed shell-like shape in the radio and the existence of nearby giant molecular clouds \citep{Erkes(1969)AJ_74_840,Huang(1986)ApJ_309_804,Tatematsu(1990)A&A_237_189}.
Such an interaction was confirmed with the detection of broad CO emission lines near the edge and the center of the remnant \citep{Koo(2001)ApJ_552_175,Byun(2006)ApJ_637_283}.
This detection also supported the suggestion that the cloud evaporation might be responsible for the enhanced thermal X-rays in the central region \citep{Leahy(1996)A&A_315_260}.

We have obtained infrared images of two specific regions of the remnant, where the shock-cloud interaction is under way.
The near- and mid-infrared band images ($\sim2-27$ \um), obtained from two infrared cameras respectively onboard the \akari{} satellite and the Palomar 5 m Hale telescope, show several shock-cloud interaction features.
We derive their brightness ratios and analyze them on the basis of the \hh{} emission model from shocked gas.
We find that their infrared properties are well described by shocked \hh{} gas with a power-law distribution of temperatures.

\section{Observations}
We observed two specific regions (``Cloud N'' and ``Cloud S'' in Fig.~\ref{fig-obs}), where it is known that slow shocks ($\la20$ \kms) propagate into clouds of density \nhh{} $\sim10^3$ \ncm{} \citep{Koo(2001)ApJ_552_175}, with two different instruments: the InfraRed Camera \citep[IRC,][]{Onaka(2007)PASJ_59_S401s} aboard the \akari{} satellite and the Wide-field InfraRed Camera \citep[WIRC,][]{Wilson(2003)inproc} at the Palomar 5 m telescope.
In this paper, we present the Cloud N data; the Cloud S data will be considered in a future paper.
The details of the observations and the data reduction on the IRC and WIRC data are described separately below.

\subsection{\akari{} IRC observations}
\akari{} was designed for both imaging and spectroscopy in the infrared \citep{Murakami(2007)PASJ_59_S369s}.
It has two scientific instruments; one, the InfraRed Camera \citep[IRC,][]{Onaka(2007)PASJ_59_S401s}, covers 2--30 \um, and the other, the Far Infrared Surveyor (FIS, \citealt{Kawada(2007)PASJ_59_S389s}), covers 50--200 \um.
\akari{} has imaged HB21 several times in either pointing or scanning modes with the IRC and FIS.
In this paper, we analyze the IRC pointing observation data.

The IRC pointing observations for Cloud N were performed at 2006 December 5--6 toward (RA, Dec) = (20:47:35.20, +51:12:07.00) in J2000.
IRC comprises three channels (NIR, MIR-S, and MIR-L), and each of these has three broad-band filters for imaging.
Among these, we employed six filters, two in each channel, for the observations \citep[IRC02 mode, see][]{Onaka(2007)PASJ_59_S401s}.
The wavelength coverage and the imaging resolution ($\Gamma$) are listed in Table~\ref{tbl-obs}, together with the pixel sizes of the detector array in each channel.

All observational data were initially processed with the IRC Imaging Pipeline \citep[v.~20070104][]{Lorente(2007)man}.
In this process, several instrumental corrections were made to the dithered images, with the corrected images then coadded.
The instrumental corrections include a dark correction, a bad-pixel masking, a distortion correction, and flat-fielding \citep[see][for detail]{Lorente(2007)man}.
We used the ``self-dark,'' a dark image measured for each individual observation, rather than the ``super-dark,'' a general dark-image periodically measured during the mission.
Astrometric information was appended to the coadded images by matching the positions of field point-sources with those of 2MASS catalog sources \citep{Skrutskie(2006)AJ_131_1163s}; the matching tolerance was 1.5 pixels.
We used the most recent conversion factor, which converts the instrumental flux (ADU) to physical flux (Jy), as noted on the \anchor{http://www.ir.isas.jaxa.jp/AKARI/Observation/DataReduction/IRC/ConversionFactor_071220.html}{\akari/IRC Data Reduction Support Page}\footnote{\url{http://www.ir.isas.jaxa.jp/AKARI/Observation/DataReduction/IRC/ConversionFactor\_071220.html}} at 2007 Dec 20.
The systematic errors ($\sim2-5\%$) of the conversion factor are included in the error estimation.

For compatibility between the images from different bands, we equalized the pixel size and the spatial resolution.
The images were resampled to have the same pixel-size ($1\arcsec\times1\arcsec$), employing the public software \anchor{http://terapix.iap.fr/rubrique.php?id_rubrique=49}{SWARP}\footnote{\url{http://terapix.iap.fr/}}, and convolved with different Gaussian kernels to have the same spatial-resolution ($\simeq7.43''$).
Then, to enhance diffuse features, we removed point sources with DAOPHOT package \citep{Stetson(1987)PASP_99_191} in IRAF; saturated point-sources were simply masked out.
The final images of Cloud N are displayed in Figure~\ref{fig-result}.

\subsection{Palomar WIRC \hh{} observations} \label{obs-wirc}
We also carried out near-infrared imaging observations of Cloud N, centered at (RA, Dec) = (20:47:36.45, +51:11:42.41) in J2000, using the 2.12 \um{} narrow-band filter for the \HH{} transition with WIRC on the Palomar 5 m Hale telescope on 2005 August 29.
WIRC is equipped with a Rockwell Science Hawaii II HgCdTe 2K infrared focal plane array, covering an $8.7'\times8.7'$ field of view with a $\sim0.25''\times0.25''$ pixel size.
We obtained 50 dithered images of 30 sec exposure. 
For the basic data reduction, we subtracted a dark and sky background from each individual dithered frame and then flat-fielded.
We then combined the dithered frames to make the final image.

The astrometry solution was obtained by matching the positions of 14 field point-sources with those of 2MASS catalog sources.
The positions were matched within $\sim0.1\arcsec$, which is smaller than the 2MASS systematic root-mean-square uncertainty of $0.15''$.
We calibrated the image by comparing the magnitude of the 14 field point-sources with the corresponding $K_s$ magnitude from the 2MASS catalog.
The correlation coefficient and the ratio between the two magnitudes was 0.9930 and $1.047\pm0.011$, respectively.
The systematic error ($\sim12\%$) of the calibration factor was included in the error estimation.
The point sources were removed with DAOPHOT as done for the IRC images, and the full-width-at-half-maximum (FWHM) of the resultant point-spread-function is $\sim1.2\arcsec$.
The final image for Cloud N is displayed in Figure~\ref{fig-result}.

\section{Results} \label{res}
The final images from the IRC and WIRC are displayed in Figure~\ref{fig-result}, together with 1420 MHz radio continuum \citep{Tatematsu(1990)A&A_237_189} and \CO{} 230.583 GHz \citep{Koo(2001)ApJ_552_175} images for reference.
The peak positions of the N1 and N2 clouds, where \cite{Koo(2001)ApJ_552_175} observed \CO{} broad molecular lines, are indicated as crosses on all images.

\subsection{Morphology} \label{res-mor}
The IRC images show different features from band to band.
Bands N3 and N4 are dominated by point sources, although N4 also shows a faint diffuse feature near the N2 cloud.
Bands S7, S11, and L15 all show similar diffuse features over the observed field.
Several filamentary feature are commonly seen: the large V-shaped feature around the N1 cloud; the short-and-horizontal features northward from the N1 cloud; the horizontal feature around the N2 cloud; the vertical feature southwestward from the N2 cloud.
The last two features are also seen in the N4 band.
Bands S7, S11, and L15 also manifest a clumpy feature at the west-end of the N2 cloud, which accompanies a cycloidal diffuse feature around it (cf.~Fig.~\ref{fig-bow}).
The L24 band shows rather extended and diffuse features, which are different from the relatively sharp features seen in the S7, S11, and L15 bands; moreover, the bright regions in the L24 band do not overlap with the N1 and N2 clouds.

The WIRC \HH{} image shows filamentary features, which are very similar with those seen in the IRC S7, S11, L15 bands; for example, the west side of the V-shaped feature around the N1 cloud, and the horizontal and vertical features near the N2 cloud.
This similarity is more recognizable when we compare the WIRC image with the RGB color image made with S7 (blue), S11 (green), and L15 (red) (see Fig.~\ref{fig-rgb}).
In this RGB image, the diffuse features commonly seen in the IRC S7, S11, L15 bands are more prominent; especially, the clumpy feature near the west-end of the N2 cloud is more identifiable.
Also, the RGB image shows that the infrared ``color'' varies from red to blue across the image.
The filamentary features seen in the WIRC image well follow the bluish features seen in the RGB image.

To quantify the similarity of the diffuse features commonly seen in the IRC S7, S11, L15 images and the WIRC \HH{} image, we measured the correlation of the features.
We sampled an area as outlined in Fig.~\ref{fig-rgb} with \emph{a yellow polygon} excluding \emph{yellow circles with a red slash}, and made scatter plots between the IRC and the WIRC images (Fig.~\ref{fig-correl}).
Each point of the scatter plots corresponds to the mean intensity of an $8''\times8''$ region.
As \emph{upper-panels} of Figure~\ref{fig-correl} show, the correlation of S7-to-S11 (0.94) and S11-to-L15 (0.95) are very high, while that of S7-L15 (0.84) is relatively low, but still high.
The correlations of the IRC images to the WIRC image vary from 0.63 to 0.78 (\emph{lower-panels} of Fig.~\ref{fig-correl}); the S7 image has the highest correlation, while the L15 image has the lowest.
This is consistent with the above description that the filamentary features seen in the \HH{} image well follow the bluish features seen in the RGB image.
The positive y-intercepts indicate that the spatially-correlated diffuse features seen in the IRC images are put on some background.
We removed these backgrounds during the intensity measurement of the diffuse feature, by subtracting the intensity of nearby regions (cf.~$\S$~\ref{res-qua} and \emph{dashed-line} boxes in Fig.~\ref{fig-rgb}).

Some of the features seen in the IRC and WIRC images contain geometrical relationships with the \CO{} and 1420 MHz radio continuum images.
The V-shaped feature overlaps with the U-shaped N1 cloud seen in the \CO{} image, and its apex is located near the N1 cloud peak position.
Westward from this apex, there is a bright region in 1420 MHz radio continuum.
The horizontal feature around the N2 cloud well matches with the N2 cloud seen in the \CO{} image, and the 1420 MHz radio image also shows a diffuse feature there.
The clumpy feature at the west end of the N2 cloud also shows an isolated clumpy feature in the \CO{} image, although it is rather vague because of the low spatial resolution of the CO image; no counter part of this feature is seen in the 1420 MHz image.
Considering the northward propagation of the remnant (cf.~Fig.~\ref{fig-obs}) and the geometrical connections with the features seen in the CO and radio continuum images, the V-shaped feature around the N1 cloud and the horizontal feature around the N2 cloud appear to represent a bow shock and a blast wave, respectively.
Similarly, the clumpy feature at the west end of the N2 cloud seems to be a shocked clump, considering the propagation direction of the remnant, together with the cycloidal diffuse feature surrounding the clumpy feature (cf.~Fig.~\ref{fig-rgb}).

\subsection{Quantitative Infrared Characteristics of the Shocked Gas} \label{res-qua}
For the analysis of the shock-cloud interaction features, identified in the previous section, we first chose three representative regions for further examination.
As the RGB image (Fig.~\ref{fig-rgb}) shows, the interaction features have different colors from region to region, which means that their infrared characteristics vary.
The three regions we chose are designated as follows: ``N1wake,'' ``N2front,'' and ``N2clump'' (see Fig.~\ref{fig-rgb}).

Then, we measured the intensity of these three regions in the IRC images by subtracting the background emission from nearby areas.
Two background regions were carefully selected for each chosen region, excluding any diffuse filamentary features (see Fig.~\ref{fig-rgb}).
To avoid any possible contamination from point sources, we also excluded these.
The measured intensities and the derived color values (S7/S11 and S11/L15) are listed in Table~\ref{tbl-result}.

The intensity in the WIRC \HH{} image was also determined by subtracting the background emission from nearby areas.
Additionally, it was extinction-corrected.
\cite{Lee(2001)inproc} estimated the foreground hydrogen nuclei column density, $N$(H)=$N$(\Hi)+2$N$(\hh), from the X-ray absorption toward the central region of HB21, and found $N$(H)$=(3.5\pm0.4)\times10^{21}$ \Ncm.
This $N$(H) corresponds to $A_{2.12 \mu m}\simeq0.22$ mag in the case of the ``Milky Way, $R_v=3.1$'' curve \citep{Weingartner(2001)ApJ_548_296,Draine(2003)ARA&A_41_241}.
The tabulated \HH{} intensities in Table~\ref{tbl-result} have been corrected to compensate for this amount of extinction.

Table~\ref{tbl-result} shows that the N1wake is the brightest in the L15 band, while the N2clump and N2front are the brightest in the S11 band.
The S7 intensity of the N1wake is below the 3-$\sigma$ limit (cf.~Fig.~\ref{fig-result}); we indicate its intensity as a 90\% upper confidence limit.
The color-color diagram of S11/L15 vs. S7/S11 (Fig.~\ref{fig-ccd}) reveals that each region has distinctive color properties: the N2front is the bluest, while the N1wake is the reddest.
This color property is also well-recognizable from the RGB map (Fig.~\ref{fig-rgb}).
The \HH{} intensity is the brightest at the N2front.
The \HH{} intensity of the N1wake is below the 3-$\sigma$ limit, again, as in S7 band; its intensity was also indicated as a 90\% upper confidence limit.
From the colors and \HH{} intensity of the selected regions, we deduce that the bluer the color is, the higher  the \HH{} intensity is.

\section{Radiation Source of the Shock-Cloud Interaction Features Observed in the \akari{} IRC Bands} \label{irc-org}
Since IRC images are broad-band images, they may contain contributions from both continuum and line emission.
Thus, to understand the observed infrared color ratios, we should first identify the radiation source which generates the shock-cloud interaction features observed in IRC S7, S11, and L15 bands.
We consider that the emission lines of \hh{} are the main source of the interaction features, based on the following arguments.

Firstly, the features seen in the IRC S7, S11, and L15 band images are similar to those seen in the Palomar WIRC \HH{} image; the correlation is about $0.63-0.78$ (cf.~$\S$\ref{res-mor} and Fig.~\ref{fig-correl}).
This strongly suggests that \hh{} emission is responsible for at least some of the features seen in the mid-infrared IRC bands.
Secondly, a recent result supports this assertion.
\cite{Neufeld(2007)ApJ_664_890} observed the shock-cloud interaction regions in several SNRs, and detected rotational lines of \hh{} and HD, together with fine-structure lines of several atoms and ions.
From their emission line maps, it was found that the emission lines show distinctive \emph{spatial} distributions, categorized into five distinct groups.
Among these groups, only the ``lines of S and \hh{}($J_{\textrm{\footnotesize{up}}}>2$)'' group spans the emission in the IRC S7, S11, L15 bands ($\sim5-20$ \um), which suggests that only the emission lines in this group can generate the \emph{highly-correlated} shock-cloud interaction features in the three IRC images (cf.~\emph{upper-panels} of Fig.~\ref{fig-correl}).
Among the emission lines belonging to this group, only \hh{} lines fall within the three IRC bands themselves.
Finally, the assertion is also theoretically supported.
\cite{Kaufman(1996)ApJ_456_611} numerically simulated slow C-type planar shocks (\vs$=10-40$ \kms), propagating into fully molecular gas of \nhh$=10^4-10^6$ \ncm.
According to their results, for a case similar to Cloud N (\vs$=20$ \kms, \nhh$=10^4$ \ncm), \hh{} lines are the dominant emission lines over the $5-20$ \um{} spectral range.

We could consider other possible sources for the emission, such as fine structure ionic lines, thermal dust continuum, Polycyclic Aromatic Hydrocarbons (PAHs) bands, and synchrotron radiation \citep[cf.][]{Reach(2006)AJ_131_1479}.
However, these do not seem likely, as we discuss below.

Within the wavelength coverage of the IRC S7, S11, L15 bands, there are two ionic lines, \Neii{} 12.8 \um{} and \Neiii{} 15.5 \um, whose emitter (\ionNeii, \ionNeiii) have been observed in the shocked regions of SNRs \citep[e.g.~][]{Raymond(1997)ApJ_482_881,Neufeld(2007)ApJ_664_890}.
Although these lines cannot generate spatially-correlated features in the three IRC band images, they may contribute to the shock-cloud interaction features seen in S11 and L15 images.
However, this does not seem likely because of their poor spatial-correlation with \hh{} emissions, well correlated with the shock-cloud interaction features seen in the IRC bands (cf.~lower panels of Fig.~\ref{fig-correl}).
%Firstly, \Neii{} and \Neiii{} show \emph{negative} spatial-correlations with \hh{}($J_{\textrm{\footnotesize{up}}}>2$) in three SNRs out of four \citep{Neufeld(2007)ApJ_664_890}.
%Secondly, the one exception shows a low correlation of $\sim0.4$, lower than the correlations of S11 and L15 to \HH{}, 0.75 and 0.63 (cf.~Fig.~\ref{fig-correl}).
\Neii{} and \Neiii{} show \emph{negative} spatial-correlations with \hh{}($J_{\textrm{\footnotesize{up}}}>2$) in three SNRs out of four, and the one exception shows a low correlation of $\sim0.4$ \citep{Neufeld(2007)ApJ_664_890}, lower than the correlations of S11 and L15 to \HH{}, 0.75 and 0.63 (cf.~Fig.~\ref{fig-correl}).
Additionally, a shock of $\ga80$ \kms{} is required for Ne to be ionized \citep{Hollenbach(1989)ApJ_342_306}, whose velocity is far higher than the velocity derived from CO observations, $\sim25$ \kms{} \citep{Koo(2001)ApJ_552_175}.
Hence, the contribution of \Neii{} and \Neiii{} to the shock-cloud interaction features seen in the three IRC bands would be negligible.

In C-shocks, the grain temperature is below $\sim$50 K \citep{Draine(1983)ApJ_264_485}, thus negligible thermal dust continuum is generated over the $5-20$ \um{} range.
Besides, below $\sim50$ K, thermal dust continuum increases toward $\ga20$ \um; hence, if significant thermal dust continuum were generated by the shocks, the IRC L24 image would also show the same shock-cloud interaction features as in the S7, S11, and L15 bands, considering the not-that-low sensitivity of the L24 band.
(The L24 sensitivity in the IRC02 observation mode is nearly half of the L15 sensitivity, \citealt{Onaka(2007)PASJ_59_S401s}.)
This is not the case.
Therefore, thermal dust continuum is not likely to generate the interaction features seen in the IRC S7, S11, and L15 images.

PAHs, thought to be formed by grain shattering in shocks \citep{Jones(1996)ApJ_469_740}, show strong and broad band features at 3.3, 6.2, 7.7, 8.6, 11.3, and 12.7 \um.
However, they have not been observed in shocks \citep{vanDishoeck(2004)ARA&A_42_119,Tielens(2008)ARA&A_46_289}.
This invisibility is attributed to the property that PAHs are heated slowly and cool efficiently, hence generate little excess emission above the background PAH emission \citep{Tielens(2008)ARA&A_46_289}.
Moreover, C-shocks of \vs$\ga25$ \kms{} are not strong enough to sputter refractory grain cores, like PAHs \citep{Draine(1983)ApJ_264_485}.
Thus, we reject PAHs as a source for the shock-cloud interaction features.

Synchrotron radiation is unlikely, as well.
\cite{Koo(2001)ApJ_552_175} derived a spectral index of $\alpha\simeq-0.28$ (flux$\propto\nu^{-\alpha}$) with intensity $I\,(\nu=325\,\textrm{MHz})\simeq0.05$ K $\simeq 160$ Jy sr$^{-1}$ at Cloud N.
If we extrapolate this intensity up to $\sim10$ \um, this corresponds to $\sim7$ Jy sr$^{-1}$.
This is far less than the observed intensity of $\sim10^5$ Jy sr$^{-1}$ (see Table~\ref{tbl-result}).

\section{Comparison to the \hh{} Infrared Emission of Shock Models} \label{shock}
Following the arguments in $\S$\ref{irc-org}, we now compare the observed infrared characteristics of the shock-cloud interaction features, to that predicted for \hh{} emission lines from various shock models.

\subsection{C-Shock: Isothermal \hh{} Gas} \label{cshock-iso}
In the N1 and N2 clouds, slow C-type shocks (\vs{} $\la20$ \kms) were observed from the CO observation \citep{Koo(2001)ApJ_552_175}.
Also, for a planar C-type shock, it is known that the shock-heated gas can be approximated as an \emph{isothermal} and isobaric slab of gas \citep{Neufeld(2006)ApJ_649_816}, in view of the \hh{} excitation diagrams predicted for such shocks \citep[e.g.][]{Kaufman(1996)ApJ_456_611,Wilgenbus(2000)A&A_356_1010}.
Hence, we first calculate the expected IRC colors from the emission lines of \emph{isothermal} \hh{} gas.

For the model calculation, we follow \citeauthor{Neufeld(2008)ApJ_678_974}'s \citeyearpar{Neufeld(2008)ApJ_678_974} assumptions.
They analyzed shock-cloud interaction features seen in \spitzer{} IRAC images based on \hh{} infrared emission lines.
We assume that the molecular gas consists of \hh{} and He only and that the density of He, \nhe, is proportional to the density of \hh, \nhh, with \nhe=0.2\nhh.
Then, we calculate  the \hh{} level populations in statistical equilibrium, only including the collisional excitation through \hh-\hh{} and \hh-He collisions.
The ortho-to-para ratio (OPR) of \hh{} is assumed to be 3.0.
Collisional excitation rates are obtained from \cite{LeBourlot(1999)MNRAS_305_802}, and other molecular data for \hh{} were obtained from the database provided by a simulation code, CLOUDY \citep[v07.02.01,][]{Ferland(1998)PASP_110_761}.
The effects of extinction are also included when modeling the IRC colors, with $N$(H)$=3.5\times10^{21}$ \Ncm, as in $\S$\ref{res-qua}.

The modeled IRC colors from isothermal \hh{} gas in statistical equilibrium are plotted as open circles ($\circ$) in Figure~\ref{fig-ccd}, together with the observed IRC colors of the selected regions.
Since pure rotational levels of \hh{} have their shorter wavelength transitions at higher energy levels, the trajectory of the expected IRC colors moves from the lower-left corner to the upper-right corner as the temperature increases.
Points for the same \nhh{} approach to the LTE value as \nhh{} increases.

Figure~\ref{fig-ccd} shows that isothermal \hh{} gas cannot explain the observed IRC colors with any combination of density and temperature, although the N1wake may possibly originate from isothermal \hh{} gas at T$\sim300$ K, since its S7/S11 value is an upper limit.
We also vary the OPR from 0.5 to 5, since the OPR is expected to be different from 3.0 in the interstellar clouds \citep{Dalgarno(1973)ApL_14_77,Flower(1984)MNRAS_209_25,Lacy(1994)ApJ_428_L69} and possibly even in shocked gas \citep{Timmermann(1998)ApJ_498_246,Wilgenbus(2000)A&A_356_1010}.
However, the observed IRC colors are not reproduced.
The predicted IRC colors at the \emph{same} temperature vary according to the adopted OPR; however, the \emph{locus} of predicted IRC colors are not much different from the OPR=3.0 case, as shown in Figure~\ref{fig-opr}.

This disagreement is consistent with the well-known fact that the \hh{} level population seen in shock-cloud interaction regions shows an ankle-like curve (cf. $\S$\ref{intro}).
The critical density of an \hh{} line transition increases as the energy level of the upper state increases \citep[cf.][]{LeBourlot(1999)MNRAS_305_802}; hence, an isothermal \hh{} gas can only produce either a \emph{straight line} (LTE) or a \emph{knee-like curve} (non-LTE) in the population diagram (cf.~Fig.~\ref{fig-pop}), neither of which are observed.

The disagreement mentioned above is understandable for the N2clump, since it may possess a shock structure similar to a bow shock, in which the shock velocity is non-uniform.
However, understanding this disagreement between the model and the observations is not straightforward for the N2front, since morphologically it looks like a planar shock from its elongated, uniform appearance, together with the northward propagation of the remnant (see Fig.~\ref{fig-rgb}).
This suggests that another process is needed to understand the IRC colors of the N2front under the planar C-shock model.
One possible explanation is the Wardle instability, which develops in C-shocks when an unstable balance between the magnetic force and the ion-neutral drag force breaks \citep{Wardle(1990)MNRAS_246_98}.
However, it has been shown that this process results in a negligible curvature in the \hh{} level population diagram \citep{Neufeld(1997)ApJ_487_283,MacLow(1997)ApJ_491_596}, contrary to the data.

\subsection{C-Shock: Power-law Distribution of \hh{} Gas Temperature} \label{cshock-pow}
The observed IRC colors are not reproduced by the planar C-shock model, in which the shocked gas is isothermal (cf. $\S$\ref{cshock-iso}).
Hence, we model an admixture of \hh{} gas with a temperature distribution.
We assume a power-law distribution for the column density based on the temperature, d$N=aT^{-b}dT$; recently, \cite{Neufeld(2008)ApJ_678_974} found that this model well describes the infrared characteristics of the shock-cloud interaction in the SNR IC443, where \hh{} emission lines are dominant.
$dN$ is the infinitesimal \hh{} column density within the temperature range ($T$,$T+dT$), $a$ is a constant related to the total \hh{} column density, \Nhh, and $b$ is the power-law index.
We adopt a temperature range for the integration as ($T_{min},T_{max})$=(100 K, 4000 K).
$T_{min}$ is set to be lower than \citeauthor{Neufeld(2008)ApJ_678_974}'s value of 300 K, since the \akari{} IRC bands (S7, S11, L15) are sensitive to lower temperature \hh{} gas than the \spitzer{} IRAC bands \cite{Neufeld(2008)ApJ_678_974} used, which only extends to 8 \um.
$T_{max}$ is set as 4000 K, since \hh{} gas rapidly dissociates above this temperature \citep[e.g.][]{LeBourlot(2002)MNRAS_332_985}.
With these variables, $a$ can be written as follows
\begin{equation} \label{eq-a}
a=\frac{\textrm{\NhhIRC}(b-1)}{T_{min}^{1-b}-T_{max}^{1-b}}
\end{equation}
, where \NhhIRC{} is a total column density of molecular hydrogen warmer than 100 K.
The effect of extinction is also included, as for the isothermal \hh{} gas case.

The modeled IRC colors for this case are plotted as filled circles ($\bullet$) in Figure~\ref{fig-ccd}.
As is evident, the observed IRC color ratios can now be reproduced with a suitable combination of \nhh{} and $b$.
Interestingly, both the N2clump and N2front have a similar $b$ value, $\sim3$, with \nhh{} of $\sim10^2-10^3$ \ncm.
The column densities are found to be \NhhIRC{} $\sim3\times10^{20}$ \Ncm.
The derived parameters from the power-law admixture model are listed in Table~\ref{tbl-par}.
For each set of parameters, the detail contributions of \hh{} line emission to IRC bands are listed in Table~\ref{tbl-cont}.
The ``Weight'' column of Table~\ref{tbl-cont} lists the weighting factor for each line in the IRC band contribution.
For example, the S11 band intensity can be calculated as follows.
\begin{equation} \label{eq-cont}
\frac{I_{S11}(\textrm{\hh})}{\textrm{MJy sr$^{-1}$}}=\frac{0.610\,I[\textrm{\hh}\,S(2)]+0.921\,I[\textrm{\hh}\,S(3)]}{10^{-4}\,\textrm{\luerg}}
\end{equation}
As Table~\ref{tbl-cont} shows, the pure-rotational \hh{} emission lines are dominant in all IRC bands.

The derived parameters are similar to those previously determined towards several SNRs, where interaction with nearby molecular clouds is occurring.
The derived \nhh{} at the N2front, \nhh=(1.8$_{-0.7}^{+2.0}$)$\times10^{3}$ \ncm, is similar to the value, derived from Large Velocity Gradient analysis of CO data for HB21, \nhh$=3.1\times10^3$ \ncm{} by \cite{Koo(2001)ApJ_552_175}.
From their \spitzer{} IRAC observation towards the SNR IC443, \cite{Neufeld(2008)ApJ_678_974} found that the IRAC color ratios were well explained with a range of power-law index $b$, 3.0--6.0.
Our $b$-values ($\sim3$) fall into near the lower end of \citeauthor{Neufeld(2008)ApJ_678_974}'s range.
Also, the column density we derive, \NhhIRC{} $\sim3\times10^{20}$ \Ncm, is similar to \Nhh{} towards shock-cloud interaction regions in four other SNRs (W44, W28, 3C391, and IC443), \Nhh{}$=(2.8-8.9)\times10^{20}$ \Ncm, which were determined from a two-temperature LTE fitting of pure-rotational \hh{} spectra with varying OPRs \citep{Neufeld(2007)ApJ_664_890}.
From these three parameters derived (\nhh, \NhhIRC, and $b$), we also determined the model predictions for the \HH{} intensities, and list these in Table~\ref{tbl-par}.
However, we note that they are a factor of $\sim17-33$ smaller than those observed (see Table~\ref{tbl-result}).
We discuss these results further in $\S$\ref{discuss}.

\subsection{Partially Dissociative J-shocks}
The partially dissociative J-shock model was proposed by \cite{Brand(1988)ApJ_334_L103} to explain the ankle-like curve of \hh{} level population (see \citealt{Burton(1989)inproca} for details on the model).
In this model, it is assumed that \hh{} gas survives the J-shock jump \citep[$\la25$ \kms{},][]{Hollenbach(1980)ApJ_241_L47}, and cools behind the shock front.
The postshock temperature covers $T\la3\times10^{4}$ K, for which the maximum corresponds to a shock velocity $\upsilon_{sh}\sim25$ \kms{} \citep{Hollenbach(1979)ApJS_41_555}.
The level populations for the surviving \hh{} gas behind the shock are determined by the cooling function, which depends on the temperature and the density of the postshock gas.
The semi-analytical model only includes \hh{} cooling (line and dissociational) and CO line cooling, and has two free parameters to adjust: the pressure ($P$) and the fractional CO abundance, \Xco$\equiv$\Nco/\Nhh.
The expected IRC colors at different ($P$, \Xco) are shown as open circles ($\circ$) in Figure~\ref{fig-other}.
Extinction is also included in modeling the IRC colors, as in $\S$\ref{cshock-iso}, adopting $N$(H)$=3.5\times10^{21}$ \Ncm.
We vary \Xco{} from 0 to $10^{-4}$ and varied $P$ from $10^3$ \ncmK{} up to $10^{11}$ \ncmK.
In comparison, \cite{Burton(1997)A&A_327_309} obtained the best fit to mid-infrared \hh{} observations toward Orion Molecular Cloud Peak 1 for $P=8\times10^{10}$ \ncmK.

Figure~\ref{fig-other} shows that none of these model parameters falls within the observed color range for any of the three regions in HB21.
The closest correspondence is for the N2front, with the parameters ($P$, \Xco)=($\sim10^{11}$ \ncmK, $\sim10^{-6}-10^{-5}$); this pressure is higher than that expected in SNRs, as we now show.
The shock velocity needs to be less than 25 \kms{} for the \hh{} to survive.
It is also known that \nhh$\sim10^3$ \ncm{} for the N2 cloud \citep{Koo(2001)ApJ_552_175}.
Hence, the postshock pressure ($\sim\rho v^2$) of the remnant would be $\sim10^8$ \ncmK, about 1000 times smaller than the derived value, $\sim10^{11}$ \ncmK, above.
\cite{Moorhouse(1991)MNRAS_253_662} and \cite{Chevalier(1999)ApJ_511_798} showed that the postshock pressure can be higher than the ambient pressure, when the radiative shell of the remnant collides with dense molecular clumps; however, it is only a factor of $\sim$20 higher.
There is also the possibility that the postshock \hh{} gas does not cool as low as a few hundred K.
In this case, the lower temperature \hh{} gas would be absent, causing the IRC colors to move towards the upper-right direction in the color-color diagram (Fig.~\ref{fig-ccd}) to bluer colors.
Thus, this could not explain the discrepancy between the model and the observations.
Overall, a partially dissociative J-shock does not seem to be a suitable model to explain the observed IRC colors.

\subsection{Non-stationary C-Shocks}
The non-stationary C-shock model describes the temporal evolution of a shock-cloud interaction after the collision.
\cite{Chieze(1998)MNRAS_295_672} showed that the non-stationary C-shock model possesses both J- and C-shock characteristics around an evolutionary time of $\sim1,000$ yr, before the steady state has been achieved.
\cite{Cesarsky(1999)A&A_348_945} showed that the modeled \hh{} level populations can be applicable to observations of the shock-cloud interaction region in the SNR IC443.
We obtain the \hh{} line intensity of five pure rotational lines, S(1)--S(5), modeled for a shock velocity $\upsilon_{sh}=25$ \kms, a preshock hydrogen nuclei density \nH=$10^4$ \ncm,  where \nH=\nh+2\nhh{}, and a preshock magnetic field $B=100\mu$G at the evolutionary times of (1.25, 1.50, 2.00, 4.00)$\times10^3$ yr, from the results of  \cite{Flower(1999)MNRAS_308_271}; this preshock density \nH{} is a little higher than that expected for Cloud N of HB21, \nhh{} $\la10^3$ \ncm{} \citep{Koo(2001)ApJ_552_175}.
The expected IRC colors for these models are overplotted as filled circles ($\bullet$) on Figure~\ref{fig-other}.
As the time elapses, the steady state C-shock is reached, and the expected color approaches that of isothermal gas; this is consistent with the result that the shocked gas behind a C-shock can be treated as an isothermal and isobaric slab of gas \citep{Neufeld(2006)ApJ_649_816}.

Figure~\ref{fig-other} shows that none of the expected colors overlaps with the observed IRC colors.
However, when considering the age of HB21, $\sim7,000$ yr, the elapsed time after the shock-cloud collision may be less than $\sim1,000$ yr.
Besides, the model parameters predict the \emph{postshock} \nH{} to be $\sim10^5$ \ncm{} \citep{Flower(1999)MNRAS_308_271}, which is 100 times higher than derived from the shocked CO gas, \nhh{} $\sim10^3$ \ncm{} \citep{Koo(2001)ApJ_552_175}.
Hence, it would be worth examining the IRC colors at earlier times ($\la1,000$ yr) and with a lower preshock \nhh{} density ($<10^4$ \ncm).
One way describing the non-stationary shock at such early times was developed by \cite{Lesaffre(2004)A&A_427_157}.
However, predictions for the \hh{} line emission from this model are not yet available, so this conjecture cannot be tested.

\section{Discussion} \label{discuss}

As shown in $\S$\ref{shock}, only the power-law admixture model can explain the observed IRC colors at the N2clump and the N2front; the other three models---the planar C-shock, the partially dissociative shock, and the non-stationary C-shock---can not simply reproduce the observed colors.
Hence, here we discuss the derived parameters for the power-law admixture model, based on two pictures of the shock-cloud interaction, in a bow and in a clumpy medium.
Additionally, we here comment on the adopted OPR value.  
Since no information on the \emph{actual} OPR is available at the moment, we assumed OPR=3.0; thus, the three parameters derived can be changed according to the adopted OPR.

\subsection{Nature of Molecular Shocks Seen in the Mid-Infrared}

\subsubsection{Bow Shock Picture}
Recently, \cite{Neufeld(2008)ApJ_678_974} showed that values they derived for the power-law index $b$, 3.0--6.0, in their power-law admixture model can be explained by paraboloidal bow shocks (i.e. a geometrical combination of planar C-shocks).
They derived a power-law index $b\sim3.8$ for a paraboloidal bow shock, when the \hh{} survives the shock (i.e. T $\la4,000$ K).
They argued that if some bow shocks have a maximum shock-velocity which is too small to fully cover the temperature range ($\la4,000$ K), then the spatially averaged value of $b$ will be higher than 3.8.
In this model, higher-$T$ \hh{} gas originates near the apexes of the bow shocks, while lower-$T$ \hh{} gas originates from the wakes of the bow shocks, since the shock velocity (the normal to the bow shock surface) decreases from the apex to the wake.

Our derived $b$-values ($\sim3$), somewhat lower than 3.8, can be understood within the bow-shock picture as follows. 
If the bow shock has a cycloidal shape (cf.~Fig.~\ref{fig-bow}), squashed enough along the axis of symmetry, the relative size of the slow shock-velocity portion of the bow will be decreased in comparison to the paraboloidal case.
This will decrease the amount of the lower-$T$ \hh{} gas, leading to a lower power-law index $b$.
The cycloidal feature surrounding the N2clump may imply this possibility, although the feature seems to locate \emph{behind} a compact clump rather than \emph{ahead} of it as the bow shock should be (see Fig.~\ref{fig-rgb}).
The N2front does not show any bow-shock-like features in the IRC and WIRC images (cf.~Fig.~\ref{fig-rgb}).
In this case, the derived $b$ value can be understood in the bow shock picture if the bow shocks are unresolved.
However, this leads to a problem associated with the size-scale for the bow shock.

The distance to HB21 is known to be $\sim0.8-1.7$ kpc \citep{Leahy(1987)MNRAS_228_907,Tatematsu(1990)A&A_237_189,Byun(2006)ApJ_637_283} and the imaging resolution of the Palomar WIRC \hh{} image is $\sim1.2\arcsec$ (cf. $\S$\ref{obs-wirc}).
This angular size means that the bow shock must be hidden within a length scale of $\sim0.01$ pc $\sim3\times10^{16}$ cm.
This length is comparable to the width of a single planar C-shock, propagating into a molecular cloud of preshock density \nH{} $\sim${} \nhh{} $\sim10^3$ \ncm{} \citep[e.g.~][]{Draine(1983)ApJ_264_485,Timmermann(1998)ApJ_498_246,Wilgenbus(2000)A&A_356_1010}.
Since a bow shock is a geometrical combination of planar shocks, this scale is too small to hide the bow shock.
In other words, the observed color ratios and the uniform-and-elongated appearance of the N2front suggest that all the \hh{} gas, with $100\la T \la4,000$ K, must be mixed within a length scale $\sim0.01$ pc, however, a bow shock, with a preshock density \nH{} $\sim10^3$ \ncm, is unable to mix \hh{} gas of such temperatures within such a length scale.

The bow shock picture has an additional difficulty in supplying the preshock \hh{} gas.
In the outflow case, for which the bow shock picture was first proposed \citep{Smith(1991)MNRAS_248_451}, the preshock \hh{} gas is supplied from the interstellar medium (ISM), which spreads far upstream; hence, the preshock \hh{} gas may be continuously supplied to the bow shock front.
On the other hand, in the SNR case, the preshock \hh{} gas is mainly supplied from the \emph{swept-up} molecular gas, which only spreads a finite distance upstream.
Hence, the amount of preshock \hh{} gas may be insufficient to develop the steady-state bow shock, in the SNR environments.

\subsubsection{Shocked Clumpy ISM Picture}
The derived model parameters for the power-law admixture model may be understood under a shocked clumpy ISM picture.
When a clump is swept up by a planar shock, a deformation of the clump is expected.
Indeed, from numerical simulations, it was shown that such a deformation bring about a \emph{cycloidal} tail when a \emph{smooth-edge} clump is swept up.
For example, \cite{Nakamura(2006)ApJS_164_477} and \cite{Shin(2008)ApJ_680_336} showed the formation of a cycloidal tail for a non-magnetized and magnetized clump, respectively.
Their collision conditions are the same; the Mach number of the shock and the density contrast of the cloud relative to the intercloud medium are both 10.
As is evident, the shapes of these tails are very similar to the cycloidal feature seen in the N2clump (see Fig.~\ref{fig-rgb}).

In view of this similarity, we conjecture the following picture.
If the conditions for the shock-cloud interactions at the N2front and the N2clump are similar \emph{except for the clump size}, the resultant excitation of the \hh{} will also be similar.
Under this assumption, this is quantified by the parameters \nhh{} and $b$ of the power-law admixture model.  
Then, two observational results for the N2clump and the N2front that are not clearly explained in the bow shock picture are now simultaneously explainable.  
First, the similar power-law index $b\sim3$ results from the similar collisional conditions.
Second, the different appearances (filamentary vs. clumpy) results from the physical size of the shocked clump.
In this picture, the size of the clumpy ISM at the N2front must be less than $\sim0.01$ pc $\sim3\times10^{16}$ cm, to be unresolved in the WIRC \HH{} image (Fig.~\ref{fig-rgb}).
Suggestively, this size is comparable to that of a shocked clump, proposed by \cite{Chevalier(1999)ApJ_511_798} to explain the high velocity molecular gas of low column density, observed in the SNR IC443.

However, this picture also has a issue related with the very linear shape of the N2front seen in the \HH{} image.
If numerous unresolved-clumps exist in the N2front, the overall appearance may be wriggly since the shock would propagate further at the less dense regions, as shown in the simulation of \cite{Patnaude(2005)ApJ_633_240}.
One possible explanation is that the size of clump is so small ($<<0.01$ pc) that the expected wriggle is also unresolved in the \HH{} image.
However, it is uncertain whether the cycloidal tail would be formed in such a small scale.

\subsection{\HH{} intensity} \label{h2s1}
The observed \emph{mid}-infrared IRC colors are well explained by a power-law admixture model of \hh{} gas temperature (cf.~Fig.~\ref{fig-ccd}).
The dominant emission lines in the IRC S7, S11, L15 bands are five pure-rotational lines, $\upsilon=0\rightarrow0$ S(1)--S(5).
In order to examine whether this model can also explain the shock seen in the \emph{near}-infrared, we compare the predicted \HH{} intensity from the derived parameters---\nhh, $b$, and \NhhIRC---to the observed intensity.
The observed and predicted \HH{} intensities are listed in Table~\ref{tbl-result} and \ref{tbl-par}, respectively.
As is evident, the observed intensities are a factor of $\sim33$ and $\sim17$ larger than the predicted intensities, for the N2clump and the N2front, respectively.

One simple explanation for this disagreement is the existence of additional \hh{} gas, whose temperature and density are both high, but whose column density is low enough to have negligible effect on the mid-infrared line intensities.
To compensate for the deficiency of the \HH{} intensity requires $N(v=1,J=3)\sim10^{11}$ \Ncm.
Hence, for example, if there is additional \hh{} gas present, in LTE with $T\sim2000$ K and \Nhh{} $\sim10^{13}$ \Ncm{}, both the mid-infrared IRC colors and the near-infrared \HH{} intensity can be understood.
A compact, unresolved shocked cloud is a candidate for producing such \hh{} emission.

Another explanation was proposed by \cite{Neufeld(2008)ApJ_678_974}, who encountered a similar disagreement between mid- and near-infrared line intensities in their observations for the SNR IC443.
They analyzed the four \spitzer{} IRAC band data (3.6, 4.5, 5.8, and 8.0 \um) with a power-law admixture model, and found that the derived \nhh{} is $\sim10^6$ \ncm{} when the 3.6 \um{} band data are \emph{not} used, while it is $\sim10^7$ \ncm{} when the 3.6 \um{} band data \emph{are} used.
The 3.6 \um{} band is dominated by the $v=1\rightarrow0$ rovibrational lines, while other bands are dominated by pure rotational lines.
Taken together, this implies a density that is higher when derived from rovibrational lines than when it is derived from pure rotational lines.

\cite{Neufeld(2008)ApJ_678_974}, however, argued that this disagreement may be caused by the omission of collisions with hydrogen atoms in the power-law admixture model.
They noted that the cross section for excitation by H is several orders of magnitude greater for rovibrational transitions than it is for pure rotational transitions \citep[see Table~1 and Figure~1 in][]{LeBourlot(1999)MNRAS_305_802}.
Hence, with only a small fraction of H, \nh{}/\nhh{}$\sim0.025$, the rovibrational transition can be dominated by collisions with H, rather than with \hh, in the temperature range 300--4000 K.
Indeed, such a fraction of \nh{}/\nhh{}$\sim0.025$, which corresponds to \nh/\nH$\sim0.012$, is expected in interstellar clouds with \nhh{} $\ga10^3$ \ncm{} \cite[see Table~1 and Figure~1 in][]{Snow(2006)ARA&A_44_367}.
Also, as \cite{Neufeld(2008)ApJ_678_974} noted, some theoretical models for shock waves \citep[e.g.][]{Wilgenbus(2000)A&A_356_1010} suggest that significant atomic hydrogen abundances can be achieved behind shocks that are fast enough to produce \hh{} at temperatures of a few thousand K.
Since we also omitted collisions with H in the power-law admixture model, the higher than expected \HH{} intensity might be understood in this way.

\section{Conclusions}
We observed the shock-cloud interaction region in the SNR HB21 at near- and mid-infrared wavelengths, with the WIRC at the Palomar telescope and the IRC aboard the \akari{} satellite.
The IRC S7, S11, and L15 band images and the WIRC \HH{} image reveal similar diffuse features, such as a blast wave, a bow shock, and a shocked clump.
We chose three representative regions---N1wake, N2clump, and N2front---and analyzed their infrared characteristics with several different shock models, on the basis that \hh{} emission lines are the radiation source of the shock-cloud interaction features.

We found that the IRC colors are well explained by an admixture model of \hh{} gas temperatures, whose infinitesimal column density varies as d$N\sim T^{-b}dT$.
Three physical parameters---\nhh, $b$, and \NhhIRC---were derived from this thermal admixture model (see Table~\ref{tbl-par}).
The derived $b$ value ($\sim3$) can be understood by a bow shock picture, whose shape is cycloidal rather than paraboloidal.
However, the bow shock interpretation has a size-scale problem for the N2front, because the observations require that the bow shock be hidden \emph{within} $\sim0.01$ pc $\sim3\times10^{16}$ cm, to be unresolved in the WIRC \HH{} image ($\Gamma\sim1.2$\arcsec); this scale is smaller than the expected size of bow shocks.
Instead of the bow shock picture, we propose a shocked clumpy ISM picture to simultaneously explain the obtained model parameters and the mid-infrared appearances.
To confirm this picture, more robust theoretical studies and observations are required.
We also compared the observed \HH{} intensity to the predicted intensity from the power-law admixture model.
The observed \HH{} intensities are a factor of $\sim17-33$ greater than the predicted ones.
This excess might be caused by either an additional component of hot, dense H2 has (which has low total column density), or through the omission of collisions with hydrogen atoms in the power-law admixture model (which results in an under-prediction of the near-IR line intensity).

\acknowledgments
This work is based on observations with \akari, a JAXA project with the participation of ESA. 
The authors thank all the members of the \akari{} project.
Also, the authors thank the referee for all the comments which make this paper clearer.
J.H.S thanks Ji-Hyun Kang for her helpful comments on the analysis.
H.G.L acknowledges the partial support of the Korea Science and Engineering Foundation to the Astrophysical Research Center for the Structure and Evolution of the Cosmos (ARCSEC)  at Sejong University
This work was supported by the Korea Science and Engineering Foundation (R01-2007-000-20336-0) and also through the KOSEF-NSERC Cooperative Program (F01-2007-000-10048-0).
This research has made use of SAOImage DS9, developed by Smithsonian Astrophysical Observatory \citep{Joye(2003)inproc}.

\bibliographystyle{D:/Work/Publication/bibtex/astronat/apj/apj}
%\bibliography{jhshinn} 
\bibliography{D:/Work/Publication/bibtex/jhshinn}

%+++++ Figures +++++
%\clearpage
\begin{figure}
\center{
}
\caption{
	Schematic description of \hh{} level population diagrams.
} \label{fig-pop}
\end{figure}

%\clearpage
\begin{figure}
\center{
}
\caption{
	Regions observed by \akari. 
	The IRC pointing-observation regions (``Cloud N'' and ``Cloud S'') are overlaid as two boxes on the 1420 MHz radio continuum image of HB21, obtained by using the synthesis telescope at the Dominian Radio Astrophysical Observatory. 
	The Palomar WIRC observations were performed toward a similar region. 
	In this paper, we present the Cloud N data.
	The 1420 MHz radio continuum image is kindly provided by T. L. Landecker.
} \label{fig-obs}
\end{figure}

%\clearpage
\begin{figure}
\center{
}
\caption{
	The \akari{} IRC and Palomar WIRC images of Cloud N, together with reference images. 
	See Table~\ref{tbl-obs} for the band definition of IRC images. 
	(\emph{top-panel}) IRC N3, N4, and S7 band images. 
	(\emph{middle-panel}) IRC S11, L15, and L24 band images. 
	(\emph{bottom-panel}) 1420 MHz radio continuum \citep{Tatematsu(1990)A&A_237_189}, \CO{} 230.583 GHz \citep{Koo(2001)ApJ_552_175}, and \HH{} 2.122 \um{} images. 
	The peak positions (``N1'' and ``N2''), where broad CO molecular lines were observed, are indicated with a `+' over all images \citep[cf.][]{Koo(2001)ApJ_552_175}. 
	Bright point-sources were simply masked out, as the white circles indicate in the upper six panels.
} \label{fig-result}
\end{figure}

%\clearpage
\begin{figure}
\center{
%\includegraphics[scale=0.4]{f4a.eps}\\
%\hspace{13mm}\includegraphics[scale=0.7]{f4b.eps}
}
\caption{
	(\emph{top}) The L15 band image of a clumpy feature at the west-end of the N2 cloud, designated as ``N2clump'' in section \ref{res-qua} (cf.~Fig.~\ref{fig-rgb}).
	(\emph{bottom}) Schematic description of the \emph{cycloidal} bowshock shape.
} \label{fig-bow}
\end{figure}

%\clearpage
\begin{figure}
\center{
}
\caption{
	The IRC RGB image (\emph{left}) and the WIRC \HH{} image (\emph{right}) of Cloud N. 
	The RGB image is composed of L15 (R), S11 (G), and S7 (B) band images, i.e. 15 \um{} + 11 \um{} + 7 \um. 
	All colors scale linearly, and fully cover the dynamic range of the diffuse features. 
	The three regions selected for study are overlaid on both images: N1wake, N2clump, and N2front. 
	The area defined by a \emph{black-or-white solid} line is the source, while the area defined by a \emph{black-or-white dashed} line is the background. 
	Circular areas around possible point sources were excluded during the intensity measurement to avoid possible contamination. 
	These areas are indicated as \emph{black-or-white circles with a red slash}. 
	Bright point-sources are masked out, and their positions are indicated by green circles with black shading.
	The area outlined with a \emph{yellow polygon} excluding \emph{yellow circles with a red slash} is the sampled area for the scatter plots between the IRC and WIRC images (cf. Fig.~\ref{fig-correl}).
} \label{fig-rgb}
\end{figure}

%\clearpage
\begin{figure}
\center{
}
\caption{
	Scatter plots between the IRC band images and the WIRC \HH{} image.
	Each point represents the mean intensity within an $8''\times8''$ region.
	The sampled area is outlined in Figure~\ref{fig-rgb} with a yellow polygon excluding yellow circles with a red slash.
	The lines plotted over the \emph{lower-panels} are the linear fitting lines.
	The units are MJy sr$^{-1}$ and \luerg{} for the IRC and \HH{} images, respectively.
} \label{fig-correl}
\end{figure}

%\clearpage
\begin{figure}
\center{
}
\caption{
	The IRC color-color diagram for Cloud N. 
	The axes represent the ratio of the intensities in the corresponding IRC bands. 
	The data points are shown by the three symbols in the legend (diamond, triangle, square). 
	The expected colors for both isothermal ($\S$\ref{cshock-iso}) and power-law-thermal ($\S$\ref{cshock-pow}) cases are indicated as \emph{open circles} ($\circ$) and \emph{filled circles} ($\bullet$), respectively.
	OPR=3.0 is assumed for both cases.
	The different types of \emph{black} lines connect points of equal \nhh{} and the LTE case. 
	The \emph{grey} solid lines connect points of equal power-law index ($b$) or equal temperature ($T$). 
	The values for the power-law index and temperature are also indicated.
} \label{fig-ccd}
\end{figure}

%\clearpage
\begin{figure}
\center{
}
\caption{
	The IRC color-color diagrams for Cloud N with the expected colors for \emph{isothermal} cases of various OPRs (cf.~$\S$\ref{cshock-iso}).
	The rest of the graph is the same with Figure~\ref{fig-ccd}.
} \label{fig-opr}
\end{figure}

%\clearpage
\begin{figure}
\center{
\includegraphics[scale=0.5,angle=90]{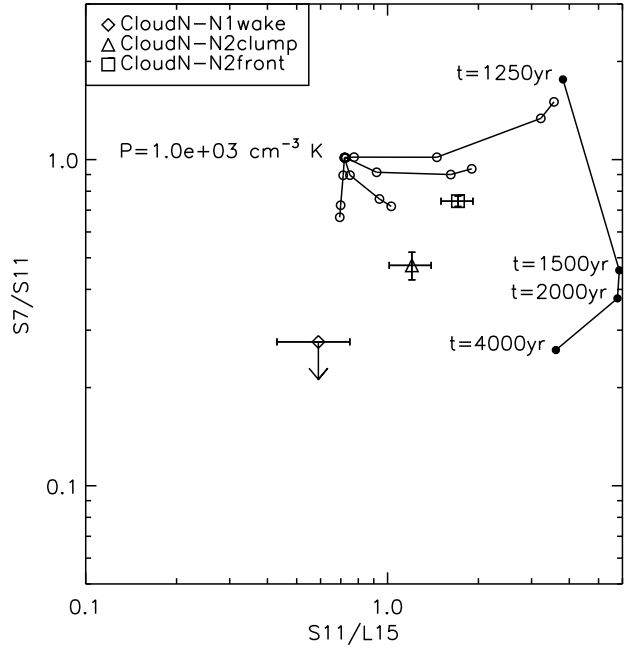}
}
\caption{
	The expected IRC colors for a partially dissociative J-shock model \citep[\emph{open circles}, $\circ$,][]{Brand(1988)ApJ_334_L103,Burton(1989)inproca} and the non-stationary shock model \citep[\emph{filled circles}, $\bullet$,][]{Chieze(1998)MNRAS_295_672,Flower(1999)MNRAS_308_271}. 
	The axes represent the ratio of the intensities in the corresponding IRC bands. 
	The data points are shown by the indicated symbols in the legend. 
	The connected \emph{open-circles} have the same fractional CO abundance to \Nhh, \Xco; the four lines correspond to \Xco$=0, 10^{-6}, 10^{-5}, 10^{-4}$, from left to right. 
	The pressure increases from $10^3$ \ncmK{} to $10^{11}$ \ncmK{}, by a factor of $10^2$, along each line. 
	Some points overlap near that for $P=10^3$ \ncmK. 
	The annotations near the \emph{filled circles} indicates the time passed in the non-stationary shock model. 
} \label{fig-other}
\end{figure}

%+++++ Tables +++++
\clearpage
\begin{deluxetable}{ccccc}
\tablewidth{0pt}
\tablecaption{Summary of the \akari{} IRC Observations \label{tbl-obs}}
\tablehead{
\colhead{Channel} &  \colhead{Filter} & \colhead{Wavelength} & \colhead{Imaging} & \colhead{Data ID} \\
& & \colhead{coverage\tablenotemark{a}} & \colhead{Resolution ($\Gamma$)}\\
\colhead{(pixel size)} &  & \colhead{(\um)} & \colhead{(FWHM, \arcsec)}
}
\startdata
NIR         &   N3 &    2.7--3.8 & 4.0 & 1400728\\
($1.46''\times1.46''$)         &   N4 &    3.6--5.3 & 4.2 & 1400728\\
MIR-S       &   S7 &     5.9--8.4 & 5.1 & 1400728\\
($2.34''\times2.34''$)       &   S11 &     8.5--13.1 & 4.8 & 1400728\\
MIR-L       &   L15 &    12.6--19.4 & 5.7 & 1400729\\
($2.51''\times2.29''$)       &   L24 &    20.3--26.5 & 6.8 & 1400729
\enddata
\tablenotetext{a}{
	Defined as where the responsivity is larger than $1/e$ of the peak for the imaging mode. 
	See \cite{Onaka(2007)PASJ_59_S401s}.
}
%\tablenotetext{\ddag}{90\% upper confidence limit.}
\end{deluxetable}

\begin{deluxetable}{ccccccc}
%\tabletypesize{\tiny}
%\tabletypesize{\scriptsize}
%\tabletypesize{\footnotesize}
\tablewidth{0pt}
\tablecaption{Results toward Cloud N \label{tbl-result}}
\tablehead{
\colhead{Region} & \colhead{S7} & \colhead{S11} & \colhead{L15} & \colhead{S7/S11} & \colhead{S11/L15} & \colhead{\HH{}\tablenotemark{a}} \\
&(MJy sr$^{-1}$) &(MJy sr$^{-1}$) &(MJy sr$^{-1}$) & & &(\luerg)
}
\startdata
N1wake                             &$<$     0.058&      0.21$\pm$0.04&      0.36$\pm$0.08&$<$      0.28&      0.59$\pm$0.16&$<$       3.1$\times10^{-7}$\\
N2clump                            &      0.33$\pm$0.02&      0.69$\pm$0.05&      0.57$\pm$0.08&      0.47$\pm$0.05&       1.2$\pm$0.2&         (2.8$\pm$0.2)$\times10^{-6}$\\
N2front                            &      0.60$\pm$0.01&      0.81$\pm$0.03&      0.47$\pm$0.05&      0.75$\pm$0.03&       1.7$\pm$0.2&         (3.8$\pm$0.1)$\times10^{-6}$
%CloudS-S1                          &      0.16$\pm$0.019&       1.0$\pm$0.012&       1.6$\pm$0.025&      0.15$\pm$0.018&      0.67$\pm$0.013&         (5.9$\pm$0.18)$\times10^{-6}$\\
\enddata
%\tablenotetext{a}{in units of \luerg}
\tablenotetext{a}{
	Extinction-corrected intensity with $N$(H)=$3.5\times10^{21}$ \Ncm. 
	See text for details.
}
\tablecomments{
	The three regions are indicated in Figure~\ref{fig-rgb}. 
	The upper limits are 90\% upper confidence limits.
}
\end{deluxetable}

\begin{deluxetable}{ccccc}
%\tabletypesize{\scriptsize}
\tablewidth{0pt}
\tablecaption{Derived Parameters for the Power-law Admixture Model\tablenotemark{a} and the predicted \HH{} intensity \label{tbl-par}}
\tablehead{
\colhead{Region} & \colhead{n(H$_2$)} & \colhead{$b$} & \colhead{\NhhIRC} & \colhead{predicted \HH{}} \\
%\colhead{Region} & \colhead{log[n(H$_2$)]} & \colhead{$b$} & \colhead{log[\Nhh]} & \colhead{predicted \hh{} S(1)} \\
&(cm$^{-3}$) & &(\Ncm) & (\luerg)
}
\startdata
CloudN-N1wake                      &$<$ 5.2$\times10^{2}$&$<$ 3.8&\nodata&\nodata\\
CloudN-N2clump                     &(5.2$_{-2.7}^{+13.6}$)$\times10^{2}$&2.9$_{-0.3}^{+0.5}$&(3.6$_{-0.6}^{+1.9}$)$\times10^{20}$&(8.6$_{-1.5}^{+2.7}$)$\times10^{-8}$\\
CloudN-N2front                     &(1.8$_{-0.7}^{+2.0}$)$\times10^{3}$&2.9$_{-0.2}^{+0.3}$&(2.5$_{-0.6}^{+1.3}$)$\times10^{20}$&(2.3$_{-0.4}^{+0.6}$)$\times10^{-7}$
%CloudS-S1                          &(3.9$_{-1.2}^{+2.1}$)$\times10^{4}$&4.2$_{-0.1}^{+0.1}$&(2.8$_{-0.5}^{+0.2}$)$\times10^{21}$&(1.5$_{-0.3}^{+0.5}$)$\times10^{-6}$\\
\enddata
\tablenotetext{a}{
	See $\S$\ref{cshock-pow} for the model description and Figure~\ref{fig-ccd} for graphical representation. 
	$b$ is the power-law index in the equation of d$N\sim T^{-b}dT$.
}
\end{deluxetable}

\begin{deluxetable}{cccccccc}
%\tabletypesize{\footnotesize}
\tablewidth{0pt}
\tablecaption{Derived Contribution of \hh{} line emission to the IRC bands \label{tbl-cont}}
\tablehead{
\colhead{Transition} & \colhead{Wavelength} & \colhead{Upper State Energy} & \colhead{IRC} & \colhead{Weight\tablenotemark{a}} & \multicolumn{2}{c}{\% Contribution} \\
\cline{6-7}
&($\mu m$) &(K) & & &N2clump &N2front
}
\startdata
H$_{2}$ $v=0-0$ $S(6)$ &     6.109 &5830 &S7 &     0.346 &4 &6\\
H$_{2}$ $v=0-0$ $S(5)$ &     6.909 &4586 &S7 &     0.530 &45 &52\\
H$_{2}$ $v=0-0$ $S(4)$ &     8.026 &3474 &S7 &     0.961 &51 &42\\
H$_{2}$ $v=0-0$ $S(3)$ &     9.665 &2504 &S11 &     0.921 &81 &84\\
H$_{2}$ $v=0-0$ $S(2)$ &    12.279 &1682 &S11 &     0.610 &19 &16\\
H$_{2}$ $v=0-0$ $S(2)$ &    12.279 &1682 &L15 &     0.023 &1 &1\\
H$_{2}$ $v=0-0$ $S(1)$ &    17.035 &1015 &L15 &     1.330 &99 &99\\

\enddata
\tablenotetext{a}{
	In units of 10$^4$ MJy sr$^{-1}$/(\luerg).
	See the text for the description.
}
\end{deluxetable}

\end{document}